\documentclass[runningheads]{llncs}

\usepackage{color}

\usepackage[T1]{fontenc}
\usepackage{graphicx}

\usepackage{amsmath}

\usepackage{xspace}
\usepackage{paralist}
\usepackage{stmaryrd}
\usepackage{amsfonts}
\usepackage{cleveref}
\usepackage{subcaption}

\usepackage{pgf}
\usepackage{tikz}
\usetikzlibrary{arrows,automata}
\usetikzlibrary{positioning}
\usetikzlibrary{shapes.geometric}
\usetikzlibrary{plotmarks}
\usetikzlibrary{patterns}
\usetikzlibrary{calc}
\usetikzlibrary{fit}
\usetikzlibrary{shapes.misc}
\usetikzlibrary{backgrounds}

\usetikzlibrary{pgfplots.groupplots}
\usetikzlibrary{shapes.multipart}

\usepackage{fancyvrb}
\usepackage{courier}
\usepackage{array}

\newcommand{\techReportAppendix}[1]{%
	\ifthenelse{\equal{\shortVersion}{true}}%
		{\cite{techReport}}%
        {Appendix~\ref{#1}}%
}%

\newcommand{\techReportAppendices}[2]{%
	\ifthenelse{\equal{\shortVersion}{true}}%
		{\cite{techReport}}%
		{Appendices~\ref{#1}--\ref{#2}}%
}%

\newcommand{\onlyTechReport}[1]{%
	\ifthenelse{\equal{\shortVersion}{true}}%
		{}%
		{#1}%
}%

\newcommand{\onlyShortVersion}[1]{%
	\ifthenelse{\equal{\shortVersion}{true}}%
		{#1}%
		{}%
}%

\newcommand{\ie}{\emph{i.e.},\xspace}
\newcommand{\eg}{\emph{e.g.},\xspace}

\definecolor{ao(english)}{rgb}{0.0, 0.5, 0.0}
\definecolor{royalblue(web)}{rgb}{0.25, 0.41, 0.88}

\newcommand{\setCommentColor}[1]{%
	\ifthenelse{\equal{#1}{bk}}%
		{\colorlet{colorVar}{red!50}}%
		{\ifthenelse{\equal{#1}{kg}}%
			{\colorlet{colorVar}{blue}}%
			{\ifthenelse{\equal{#1}{mg}}%
				{\colorlet{colorVar}{ao(english)}}%
			{\ifthenelse{\equal{#1}{jr}}%
				{\colorlet{colorVar}{magenta}}%
				{\ifthenelse{\equal{#1}{zw}}%
					{\colorlet{colorVar}{orange}}%
					{\ifthenelse{\equal{#1}{gm}}%
						{\colorlet{colorVar}{cyan}}
						{}
					}%
				}%
			}%
		}%
	}%
}

\newcommand{\commentAuthor}[1]{%
	\ifthenelse{\equal{#1}{bk}}%
		{Boris:\ }%
		{\ifthenelse{\equal{#1}{kg}}%
			{Klaus:\ }%
			{\ifthenelse{\equal{#1}{mg}}%
				{Marco:\ }%
			{\ifthenelse{\equal{#1}{jr}}%
				{Jan:\ }%
				{\ifthenelse{\equal{#1}{zw}}%
					{Zilong:\ }%
					{\ifthenelse{\equal{#1}{gm}}%
						{Gideon:\ }%
						{}
					}%
				}%
			}%
		}%
	}%
}

\definecolor{codegray}{gray}{0.95}

\newcommand{\powerset}[1]{2^{#1}}

\newcommand{\liT}{\ensuremath{\mathtt{li}}}
\newcommand{\divT}{\ensuremath{\mathtt{div}}}

\newcommand{\myvalue}[1]{\llbracket#1\rrbracket}

\newcommand{\archStates}{\textsc{ArchState}}

\newcommand{\uarchStates}{\mu\textsc{ArchState}}
\newcommand{\archStep}{\textsc{Isa}}
\newcommand{\archEval}{\archStep^*}
\newcommand{\implStates}{\textsc{ImplState}}

\newcommand{\implStep}{\textsc{Impl}}
\newcommand{\implEval}{\implStep^*}
\newcommand{\implEvalFilter}[1]{\implStep^*| #1}
\newcommand{\archProj}[1]{{#1}_{\textsc{Arch}}}
\newcommand{\uarchProj}[1]{{#1}_{\textsc{µArch}}}
\newcommand{\ctrObs}{\textsc{CtrObs}}
\newcommand{\atkObs}{\textsc{AtkObs}}
\newcommand{\ctrsat}[2]{\arch, \uarch \vdash \textsc{CtrSat}(#1,#2)}
\renewcommand{\ctrsat}[2]{#2(\uarch) \models #1(\arch)}
\newcommand{\ctrnonsat}[2]{\arch, \uarch \not\vdash \textsc{CtrSat}(#1,#2)}
\renewcommand{\ctrnonsat}[2]{#2(\uarch) \not\models #1(\arch)}
\newcommand{\mctrsat}[2]{\uarch \vdash \textsc{CtrSat}(#1,#2)}
\renewcommand{\mctrsat}[2]{#2(\uarch) \models #1(\uarch|_{\phi})}

\newcommand{\programs}{\textsc{Program}}
\newcommand{\prog}{\textsc{Prog}}

\newcommand{\nonint}{\ensuremath{P \vdash NI(\pi, \atk(\uarch))}}
\newcommand{\nonintcontract}{\ensuremath{P \vdash NI(\pi, \contract_S(\arch))}}

\newcommand{\kywd}[1]{\mathbf{#1}}

\newcommand{\startObsKywd}[1]{\obsKywd{start}}
\newcommand{\commitObsKywd}[1]{\obsKywd{commit}} 
\newcommand{\rollbackObsKywd}[1]{\obsKywd{rollback}}

\newcommand{\commitObs}[1]{\commitObsKywd{}} %
\newcommand{\rollbackObs}[1]{\rollbackObsKywd{}} %

\newcommand{\fetch}[1]{
\ifthenelse{\equal{#1}{}}{\kywd{fetch}}{\kywd{fetch}}%
}

\definecolor{orange}{rgb}{0.8, 0.47, 0.196}

\usepackage{letltxmacro}
\newcommand*{\SavedLstInline}{}
\LetLtxMacro\SavedLstInline\lstinline
\DeclareRobustCommand*{\lstinline}{%
  \ifmmode
    \let\SavedBGroup\bgroup
    \def\bgroup{%
      \let\bgroup\SavedBGroup
      \hbox\bgroup
    }%
  \fi
  \SavedLstInline
}

\newcommand{\compose}[2]{#1 \triangleright #2}
\renewcommand{\compose}[2]{#1(#2)}		%

\newcommand{\arch}{\archStep} %
\newcommand{\uarch}{\implStep} %

\newcommand{\isasat}[3]{#1 \equiv_{#3} #2}
\renewcommand{\isasat}[3]{#1 \vdash_{#3} #2}

\newcommand{\atk}{\textsc{Atk}}
\newcommand{\contract}{\textsl{\textsc{LC}}}

\begin{document}
\title{Foundations of\\Hardware-Software Leakage Contracts}
\author{Jan Reineke\orcidID{0000-0002-3459-2214}}
\authorrunning{J. Reineke}
\institute{Saarland University, Saarland Informatics Campus, Saarbrücken, Germany\\
\email{reineke@cs.uni-saarland.de}}
\maketitle              %
\begin{abstract}

Microarchitectural attacks such as Spectre and Meltdown undermine security by exploiting software-visible side effects of microarchitectural optimizations like caches and speculative execution. 
To use modern hardware securely, programmers must take into account the security implications of these optimizations. 
Yet instruction set architectures (ISAs)---the conventional abstraction layer between hardware and software---are ill-suited for secure programming: by design, they abstract from microarchitectural details and therefore fail to reflect their security consequences.

A new security abstraction is emerging to fill this void: \emph{hardware-software leakage contracts}. 
To support the principled development of secure software on modern hardware such contracts characterize side-channel leakage at the ISA level. 
Thereby they enable programmers and compilers to generate code that is provably resistant to side-channel attacks---without having to explicitly reason about the microarchitecture. 
On the other hand, they provide a specification for hardware designers to ensure that microarchitectures do not introduce unintended and undocumented side channels.
This paper introduces the foundations of this emerging topic, discussing recent contributions and unresolved challenges along the way.\looseness=-1

\end{abstract}
\section{Introduction}\label{sec:introduction}

Modern microarchitectures implement a growing number of optimizations to increase performance. 
Today even simple processors in embedded systems often employ caches to hide the latency of main memory.
General-purpose processors in personal computers and data centers employ multi-level caches, deep pipelines, and various forms of speculation to reduce latency and to increase instruction throughput.

Microarchitectural attacks~\cite{Aciicmez07,Ristenpart:prime+probe+llc,Yarom14,Liu15,Yarom2017,spectre2019,ret2spec2018,stecklina2018lazyfp,Aldaya2019,fallout2019,zombieload2019,ridl2019,Weber0NSR21,GerlachWZS23,PurnalBPV23} exploit the fact that these features may have software-visible side effects.
In particular, instructions may have vastly different execution latencies depending on the context they are executed in.
For example, memory instructions can be served in a few cycles from the cache but may take hundreds of cycles if they need to access main memory.
Measuring latencies of instruction or of entire programs thus allows attackers to gain information about the execution context, which may in turn depend on secret data.

Defending against microarchitectural attacks is challenging since microarchitectural optimizations are ``invisible'' at the instruction set architecture (ISA) level, which is the traditional hardware-software interface.
\emph{Hardware-software leakage contracts}~\cite{contracts2021} (short: leakage contracts) aim to fill this void by specifying at the ISA-level what information a program may leak. 
In this way they may serve as a rigorous foundation for writing programs that are resistant against microarchitectural attacks.

\newcommand{\mycaption}{\vspace{-3mm}
        \caption{}
        \vspace{5mm}}
\begin{figure}[t]
    \hfill~
    \begin{subfigure}[t]{0.31\textwidth}
        \begin{Verbatim}[commandchars=\#\[\]]
#textbf[if] (secret)
    wait 1 second
#textbf[else] 
    wait 100 seconds
#textbf[return] 42
        \end{Verbatim}
        \mycaption\label{fig:a}
    \end{subfigure}
    ~\hfill
    \begin{subfigure}[t]{0.34\textwidth}
        \begin{Verbatim}[commandchars=\#\[\]]

#textbf[return] 
    input * secret


        \end{Verbatim}
        \mycaption\label{fig:b}
    \end{subfigure}
    \begin{subfigure}[t]{0.31\textwidth}
        \begin{Verbatim}[commandchars=\#\[\]]
#textbf[if] (secret)
    #textbf[return] 23
#textbf[else] 
    #textbf[return] 42

\end{Verbatim}
        \mycaption\label{fig:c}
    \end{subfigure}
    \begin{subfigure}[t]{0.1\textwidth}
        ~
    \end{subfigure}
   \begin{subfigure}[t]{0.4\textwidth}
        \begin{Verbatim}[commandchars=\#\{\}]

nirvana = Mem[secret]
#textbf{return} Mem[input]

\end{Verbatim}
        \mycaption\label{fig:d}
    \end{subfigure}
    \begin{subfigure}[t]{0.3\textwidth}
        \begin{Verbatim}[commandchars=\#\{\}]
#textbf{assert} (input >= A.size)
#textbf{if} (input < A.size)   
    y = A[input]   
    z = B[y]
\end{Verbatim}
        \mycaption\label{fig:e}
    \end{subfigure} 
\caption{Example programs that may or may not leak secrets via their timing.}\label{fig:examples}
\end{figure}

Consider the pseudo-code example programs in \Cref{fig:examples}.
Let us assume an attacker who can control the \verb~input~ and observe a program's execution time.
Which of these programs leak information about secrets to this attacker?

The program in \Cref{fig:a} certainly leaks the value of the secret variable via its timing, independently of the underlying microarchitecture.
This example illustrates that side-channel security cannot be achieved purely in hardware.

For the program in \Cref{fig:b} it depends.
On some processors, the multiplication instruction has operand-dependent timing, e.g. due to a fast-path implementation of multiplication by zero.
On such processors, the timing of the program in \Cref{fig:b} will give away some information about \verb~secret~.
On other processors, the same program may have constant execution time.

For the program in \Cref{fig:c} it again depends.
Assuming that the conditional statement is implemented via a conditional branch instruction, the program will leak the branch outcome---and thus the secret---on most in-order pipelined processors.
There, the ``branch taken'' case will take a few cycles longer as the processor initially continues fetching the instructions immediately following the branch instruction, switching to the correct branch target only once the branch condition has been evaluated.
Most processors also employ instruction caches. 
With instruction caches, the two program paths may touch a different number of memory blocks and thus have secret-dependent execution time due to a varying number of cache misses.

On simple embedded processors without caches or Dynamic Random-Access Memory (DRAM)-based main memory the program in \Cref{fig:d} may have constant execution time.
In the presence of a data cache, on the other hand, the program may be used to leak some bits of the value of \verb~secret~.
Assuming an initially empty cache, the access to \verb~Mem[input]~ will result in a cache hit if and only if it accesses the same memory block as the first memory access to \verb~Mem[secret]~.
Thus, depending on the exact cache geometry a subset of the bits of \verb~secret~ leak.
E.g. on a byte-addressable system with 32-byte cache lines, the 5 least significant bits determine the offset within a memory block, while all other bits determine the memory block and may thus be inferred from the execution time of the program by varying the \verb~input~ and measuring the execution time.
Similarly, the program's execution time may leak information about the secret due to history-dependent DRAM latencies.\looseness=-1 %

Unlike the previous examples, the program in \Cref{fig:e} has no explicit secrets.
For the sake of the example, let's assume that the array bounds and the data within arrays \verb~A~ and \verb~B~ is public while all out-of-bounds data is secret.
On non-speculative processors the program in \Cref{fig:e} either fails on the assertion or it branches to the end of the program.
Thus without speculation there's no leakage of secret data.
On processors implementing speculative execution without Spectre countermeasures, however, the program may leak out-of-bounds data of array \verb~A~:
Assume that \verb~input >= A.size~.
Then, the processor may nevertheless speculatively execute the two accesses to \verb~A[input]~ and \verb~B[y]~, later rolling back their architectural effects.
The first access is out-of-bounds and thus brings secret data into variable \verb!y!, and the second access encodes this secret value in the cache state, as the address of the access is secret dependent.
This second access also makes the execution time of the program itself secret dependent as it may result in a hit or a miss depending on the value of \verb~A[input]~ and the initial cache state.
The program is a simple variation of Spectre V1~\cite{spectre2019}.

From the above examples we can draw several conclusions:
\begin{asparaenum}
    \item Side-channel vulnerabilities are not a matter that can be fully addressed by clever microarchitecture design. The program in \Cref{fig:a} leaks the secret on any reasonable microarchitecture.
    \item The same program may be vulnerable on one microarchitecture and secure on another (\Crefrange{fig:b}{fig:e}). Furthermore, depending on the microarchitecture, data may leak even if the program's execution logically does not depend upon it, as the Spectre~V1 example in \Cref{fig:e} shows.
\end{asparaenum}

\begin{figure}[t]
\begin{center}
\begin{subfigure}[t]{0.49\textwidth}
    \includegraphics[height=0.4\textwidth]{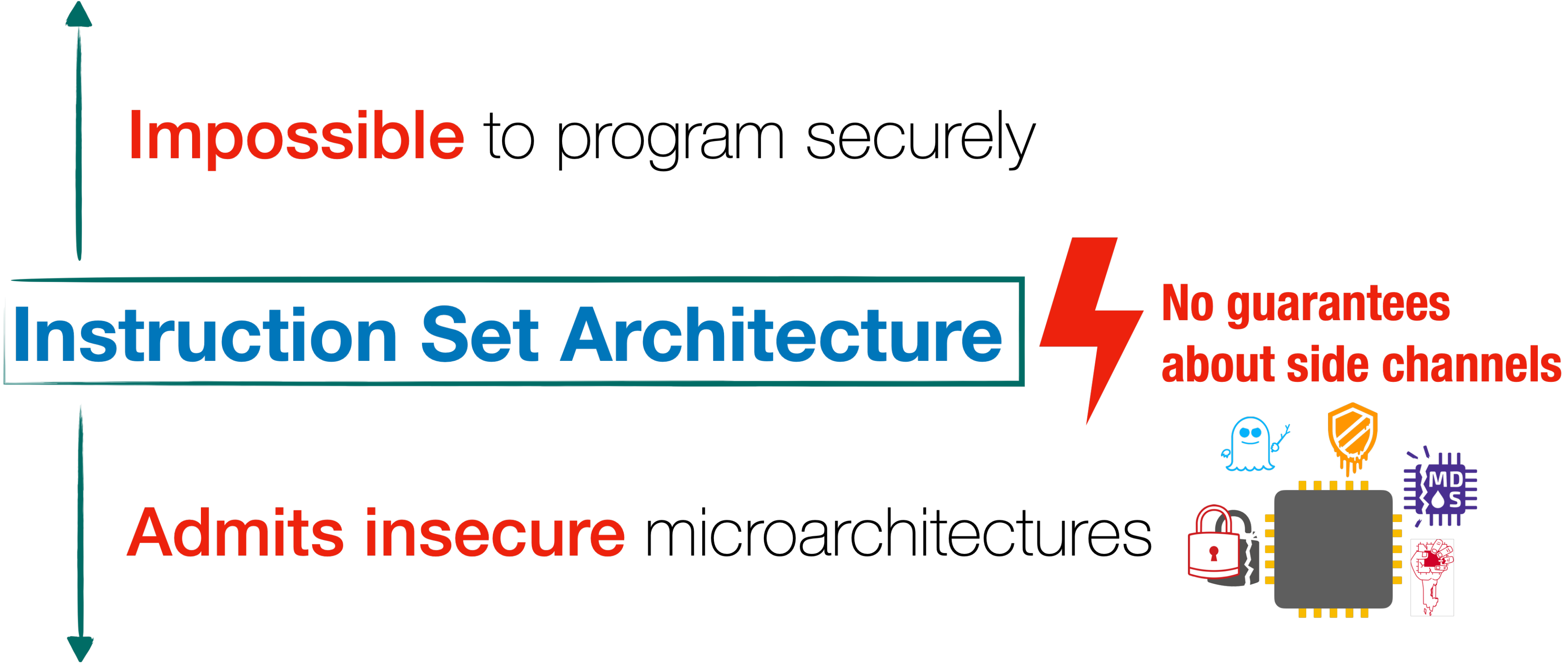}
    \mycaption
\end{subfigure}
\begin{subfigure}[t]{0.49\textwidth}
    \includegraphics[height=0.4\textwidth]{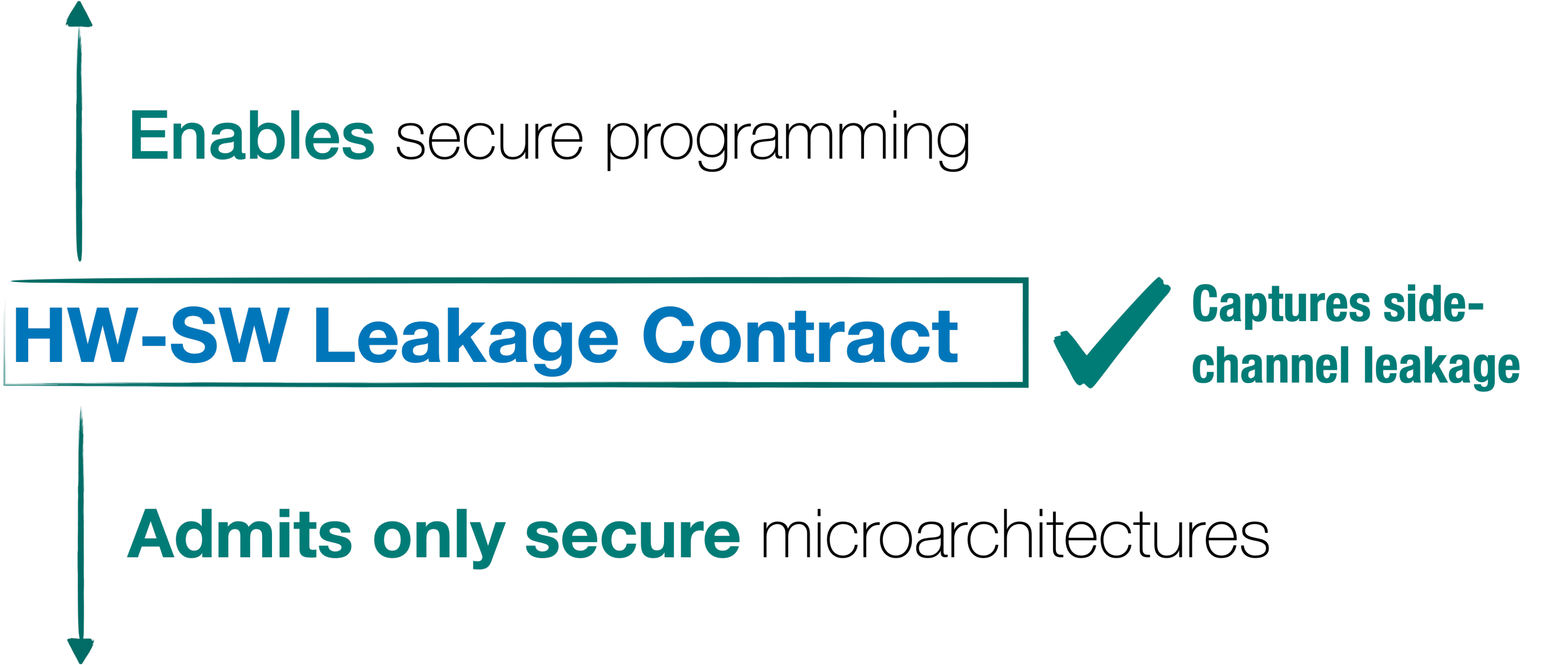}
    \vspace{1mm}
    \mycaption
    \end{subfigure}
\caption{Leakage contracts establish security at the ISA level.}\label{fig:vision}
\end{center}
\end{figure}

Due to the absence of guarantees about side-channel leakage the ISA level, programmers need to reason directly about the underlying microarchitecture to write side-channel-secure code. 
However, vendors are not releasing specifications about the microarchitecture at a sufficient level of detail to guarantee security. 
Furthermore, future microarchitects may design optimizations that break current programs' security by introducing new sources of leakage without breaking any guarantees provided by the ISA.
As illustrated in \Cref{fig:vision}, leakage contracts address both of these issues.
They specify potential information leakage due to side channels at the ISA level. 
This enables programmers and compiler designers to reason about side-channel security without having to explicitly consider microarchitectural features.
Leakage contracts also act as a specification that constrains the design of future microarchitectures, preventing the unintended and undocumented introduction of new vulnerabilities.

Consider the following informal contract in plain language:\\
``For every multiplication the second operand leaks and for every memory access the memory address leaks.''

Interpreting a program instruction-by-instruction and evaluating the contract in each step yields a trace of contract observations.
For example, for any initial state, the program in \Cref{fig:b} yields a trace consisting of a single observation that leaks the value of \verb~secret~.
The program in \Cref{fig:d} yields traces of length two, leaking \verb~secret~ and \verb~input~.
If \verb~wait x seconds~ is implemented using a busy-wait loop, then \verb~wait 1 second~ and \verb~wait 100 seconds~ execute different numbers of instructions.
As a result, their contract traces differ in length and thus implicitly leak the value of \verb~secret~.
Thus these three programs are vulnerable under the contract above, as they leak information about secrets.
On the other hand, the programs in \Cref{fig:c,fig:e} are secure as they yield traces that are independent of any secrets.

We discuss \emph{secure programming}, \ie what it means for a program to be secure under a contract, as well as techniques to verify and enforce this property in \Cref{sec:programming}.

Secure programming is only meaningful if the microarchitecture that a program is executed on satisfies the contract. 
A contract is \emph{sound} for a given microarchitecture, or, equivalently, the microarchitecture satisfies the contract, if the contract captures \emph{all} leaks in the microarchitecture, \ie any two program executions that are distinguishable by an attacker must result in different contract traces.
We formally define contracts and contract satisfaction in \Cref{sec:contracts}.
There, we also discuss techniques to verify that a given microarchitecture satisfies a contract and to synthesize contracts for a given microarchitecture.
To prepare for this, we introduce architectures and microarchitectures in \Cref{isamuarch}.

\section{Architectures and Microarchitectures}\label{isamuarch}

Our formalization of architectures and microarchitectures follows prior work in~\cite{wang2025synthesis}, which in turn builds on \cite{Mohr24,contracts2021}.
\subsubsection{Architectures}\label{sec:arch}

We model architectures as state machines that define how the execution progresses through a sequence of architectural states. 
Each transition in such a sequence corresponds to the execution of a single instruction.
Formally, an \emph{architecture} is a pair $(\archStates, \archStep)$ where $\archStates$ is a set of architectural states and $\archStep: \archStates \to \archStates$ is a transition function that maps each state $\sigma \in \archStates$ to its successor $\archStep(\sigma)$, modeling the execution of one instruction.
In most modern architectures, architectural states are composed of the state of the memory and of the architectural registers, including the program counter.
As part of the state of the memory, the program to be executed is also part of the architectural state and we assume that it can be extracted via a function $\prog : \archStates \rightarrow \programs$. 

We denote by $\archEval(\sigma)$ the sequence of states reached from $\sigma$ by successive applications of $\archStep$, \ie $\archEval(\sigma) = \sigma_0, \sigma_1, \sigma_2, ...$ where $\sigma_0 = \sigma$ and $\sigma_{i+1} = \archStep(\sigma_i)$ for all $i \geq 0$.

\begin{example}[A simple instruction set~\cite{wang2025synthesis}]\label{example:isa}
Consider a simple instruction set $\arch$ that supports loading immediate
values into registers and performing divisions between them.
The architectural state consists of $n$ 32-bit registers, interpreted as
signed integers, which we identify using $\mathtt{R1}, \ldots, \mathtt{Rn}$.
It also contains a dedicated register $\mathtt{PC}$ holding the program counter. 
The instruction set consists of two instructions: 
\begin{asparaenum}
  \item The \textbf{load immediate} instruction $\liT\ \mathtt{RD}, \mathtt{imm}$ sets the
  value of the destination register $\mathtt{RD}$ to $\mathtt{imm}$.
  $\mathtt{RD}$ is a register identifier in $\mathtt{R1}, \ldots, \mathtt{Rn}$
  and $\mathtt{imm}$ is a 32-bit value.
  \item The \textbf{division} instruction $\divT\ \mathtt{RD}, \mathtt{RS1}, \mathtt{RS2}$ divides register $\mathtt{RS1}$ by register $\mathtt{RS2}$ and stores the result in register $\mathtt{RD}$.
  $\mathtt{RD}, \mathtt{RS1},$ and $\mathtt{RS2}$ are register identifiers in
  $\mathtt{R1}, \ldots, \mathtt{Rn}$.
\end{asparaenum}
\end{example}

Most of today's commercial processors implement x86 or ARM instruction sets.
Significant efforts have been undertaken to formalize the ARM instruction sets~\cite{DBLP:conf/fmcad/Reid16,DBLP:conf/cav/ReidCDGHKPSVZ16}.
In particular, ARM created the Architecture Specification Language (ASL) to formally specify its architectures and they released a formal model of the ARMv8-A architecture~\cite{reid2017arm_v8a_xml_release} in a machine-readable and executable format.
The official x86 specification is published by Intel and AMD in their processor manuals~\cite{intel-sdm,amd-apm}.
Compared with ARM, however, relying on pseudo code and prose, the x86 specification is informal and thus less amenable to formal analysis.
There have been various academic efforts to formalize x86~\cite{DBLP:phd/dnb/Degenbaev12,DBLP:conf/fmcad/GoelHKG14,DBLP:conf/pldi/HeuleS0A16,DBLP:books/sp/17/GoelHK17,DBLP:conf/sigsoft/HasabnisS16,DBLP:conf/pldi/Dasgupta0KAR19,DBLP:conf/cpp/RoessleVR19}, but none has been adopted by Intel or AMD.
Formal ISA specifications enable formal verification of microarchitectures~\cite{DBLP:conf/cav/ReidCDGHKPSVZ16}, formally reasoning about software~\cite{DBLP:conf/pldi/SammlerHL0PD0S22} and verified code compilation~\cite{DBLP:journals/cacm/Leroy09}.

Sail~\cite{DBLP:journals/pacmpl/ArmstrongBCRGNM19} is a domain-specific language for defining ISA semantics. 
It has been used to formalize multiple ISAs including ARM, \mbox{RISC-V}, and CHERI~\cite{DBLP:conf/isca/WoodruffWCMADLNNR14}. 
\mbox{RISC-V}~\cite{Waterman:EECS-2014-54} is an open ISA that has been gaining traction in both academia and industry.
As an open ISA, RISC-V attracted a growing number of open-source implementations, which are a promising target for research on hardware-software leakage contracts.

\subsubsection{Microarchitectures}\label{sec:uarch}
We model microarchitectures as state machines that define how the processor's state evolves at cycle granularity.
Thus, a \emph{microarchitecture} is a pair $(\implStates, \implStep)$ where
 $\implStates = \archStates \times \uarchStates$ is the set of microarchitectural states %
 and $\implStep: \implStates \to \implStates$ is a transition function that maps each state $\sigma \in \implStates$ to its successor $\implStep(\sigma)$, modeling the execution of one cycle.

Microarchitectural states $\sigma \in \implStates$ consist of an architectural part $\archProj{\sigma} \in \archStates$, directly corresponding to the ISA state, and a microarchitectural part $\uarchProj{\sigma} \in \uarchStates$ modeling the state of microarchitectural components, such as caches and predictors.
Just like  $\archEval(\sigma)$,  $\implEval(\sigma)$ denotes the sequence of states reached from $\sigma$ by successive applications of $\implStep$.
Given a predicate $\phi$ over $\implStates$, $\implEvalFilter{\phi}(\sigma)$ denotes the subsequence of states from $\implEval(\sigma)$ that satisfy $\phi$.
Finally, given a sequence of microarchitectural states $\tau := \tau^0, \tau^1, \tau^2 ...$ we denote by $\archProj{\tau}$ the sequence of architectural states obtained by applying the architectural projection to all elements of $\tau$, \ie $\archProj{\tau} := \archProj{\tau}^0, \archProj{\tau}^1, \archProj{\tau}^2 ...$

For simplicity, we  refer to a microarchitecture only using its transition function $\implStep$ rather than the tuple $(\implStates, \implStep)$.
Similarly, we refer to an architecture $(\archStates, \archStep)$ simply using its transition function~$\archStep$.\looseness=-1

\begin{example}[A simple processor~\cite{wang2025synthesis}]\label{example:processor}%
Consider a two-stage pipelined implementation $\uarch$ of $\arch$.
The \emph{fetch+decode stage} takes one cycle.
It loads the current instruction at the address in $\mathtt{PC}$, decodes it into instruction type and operands, and increments the program counter.

The \emph{execute stage} works as follows:
\begin{asparaitem}
  \item Load immediate instructions $\liT\ \mathtt{RD}, \mathtt{imm}$ are executed in a single
  cycle by writing the  value $\mathtt{imm}$ to the destination register
  $\mathtt{RD}$.
  \item Division instructions $\divT\ \mathtt{RD}, \mathtt{RS1}, \mathtt{RS2}$ are handled by a dedicated division unit. 
  The unit takes 32 cycles for all divisions except when $\mathtt{RS2}$ is $0$ or $1$, in which case it takes 1 cycle only. 
\end{asparaitem}

Whenever the execute stage takes more than one cycle, the fetch+decode stage gets stalled.
\end{example}

Of course, commercial microarchitectures are significantly more complex than the simple two-stage pipeline from \Cref{example:processor}.
They employ multi-level caches, out-of-order execution, and various forms of speculation to increase performance.
All of these features have been shown to be exploitable via microarchitectural attacks~\cite{Aciicmez07,Ristenpart:prime+probe+llc,Yarom14,Liu15,Yarom2017,spectre2019,ret2spec2018,stecklina2018lazyfp,Aldaya2019,fallout2019,zombieload2019,ridl2019,Weber0NSR21,GerlachWZS23,PurnalBPV23}.
Unfortunately, commercial microarchitectures are typically closed source and thus not amenable to formal analysis.
The growing body of RISC-V processors~\cite{darkriscv,sodor,cva6,rocket-chip,ibex,Celio:EECS-2015-167} is attractive for research on leakage contracts as these designs are open source and come in a range of complexities, from simple in-order designs to complex out-of-order designs with speculation.

\subsubsection{Functional Correctness}\label{sec:functionalcorrectness}

To correctly implement an architecture $\arch$, a microarchitecture $\uarch$ needs to transition through microarchitectural states in a manner consistent with $\arch$.
For in-order scalar processors, \ie processors executing and retiring at most one instruction per cycle, this is adequately captured by the \emph{ISA compliance} notion from Wang et al.~\cite{Wang23}.
This notion relies on a predicate $\phi : \archStates \rightarrow \mathbb{B}$ that indicates when $\uarch$ retires instructions.
They call such a predicate a \emph{retirement predicate}.

In a pipelined processor, $\phi$ will hold when the oldest in-flight instruction reaches the write-back stage and its architectural effects become visible. 
At the start of execution, $\phi$ will remain false until the pipeline has been sufficiently filled and the first instruction reaches the write-back stage. 
In the steady state, once the pipeline is full, $\phi$ will hold in every cycle in which an instruction retires—typically one per cycle for an in-order scalar pipeline.
However, $\phi$ may not hold in every cycle, for instance when the pipeline is stalled due to hazards, cache misses, or other long-latency operations that prevent retirement.

Remember that $\implEvalFilter{\phi}(\sigma)$ denotes the subsequence of states from $\implEval(\sigma)$ that satisfy $\phi$.
Thus $\archProj{(\implEvalFilter{\phi}(\sigma))}$ is the sequence of architectural states observed at instruction retirements during the execution starting from microarchitectural state $\sigma$.

We say that a microarchitecture $\uarch$ correctly implements an architecture $\arch$ if the changes in $\uarch$’s architectural state can be mapped to executions of $\arch$ using $\phi$.
This is captured formally by the following definition:

\begin{definition}[ISA compliance~\cite{Wang23}]\label{def:isa-compliance}
Microarchitecture $\uarch$ correctly implements architecture $\arch$ with respect to a retirement predicate $\phi$ over $\archStates$, denoted by $\isasat{\uarch}{\arch}{\phi}$, if for all states $\sigma \in \implStates$:
\begin{enumerate}[(a)]
\item $\archProj{(\implEvalFilter{\phi}(\sigma))} = \archEval( \archProj{\sigma} )$, \ie all architectural states witnessed by $\phi$ agree with $\arch$, and 
\item $\archProj{(\implStep^i(\sigma))} = \archProj{(\implStep^{i-1}(\sigma))}$ whenever $\archProj{(\implStep^i(\sigma))} \not\models \phi$, \ie  no architectural changes beyond those witnessed by $\phi$.\looseness=-1 
\end{enumerate}
\end{definition}

While we are mainly concerned with side-channel leakage rather than with functional correctness  it turns out that the two notions are tightly coupled.
For non-trivial leakage contracts, it is practically impossible to construct a microarchitecture that satisfies the contract without implementing the architecture correctly.
We will discuss how security and functional correctness can be decoupled in \Cref{sec:contracts}, which allows testing and verification to focus either on security or functional correctness.

\section{Hardware-Software Leakage Contracts}\label{sec:contracts}

Following~\cite{Mohr24,wang2025synthesis} we define leakage contracts not as monolithic objects, but as sets of \emph{contract atoms}, which correspond to basic units of leakage.

\subsubsection{Contract Atoms~\cite{Mohr24,wang2025synthesis}}
Atoms characterize potential instruction-level leaks and are composed of two parts: an \emph{applicability predicate} and a \emph{leakage function}. %
The former determines whether a contract atom is applicable in a given architectural state, while the latter specifies what information from the architectural state is exposed when the atom is applicable.

Formally, a \emph{contract atom} $A$ is a pair $A = (\pi_A, \phi_A)$, where the applicability predicate $\pi_A$ is a predicate over $\archStates$ and the leakage function $\phi_A$ is a function $\phi_A : \archStates \to \ctrObs$, where $\ctrObs$ denotes the set of possible contract observations.

One possible way to systematically derive contract atoms for a given ISA is to create one atom for each operand of every instruction type. The applicability predicate of an atom identifies the instruction type it applies to, while the leakage function specifies which part of the instruction (such as a particular operand) is leaked.
To make this concrete, \Cref{example:atoms} shows contract atoms constructed in this manner for the ISA from \Cref{example:isa}.

\begin{example}[Contract atoms]\label{example:atoms}
The ISA from \Cref{example:isa} supports two types of instructions: \liT\ and \divT.
Three possible contract atoms for these instructions are:\looseness=-1
\begin{compactitem}
    \item $A_1 = (\liT, \langle\mathtt{``imm"}, \myvalue{\mathtt{imm}}\rangle)$, where the applicability predicate $\liT$ holds whenever the next instruction is a load immediate instruction, and the leakage function $\langle\mathtt{``imm"}, \myvalue{\mathtt{imm}}\rangle$ leaks the immediate value $\myvalue{\mathtt{imm}}$ of the instruction together with a leakage identifier $\mathtt{``imm"}$.
    \item $A_2 = (\divT, \langle\mathtt{``Reg[RS1]"}, \myvalue{\mathtt{Reg[RS1]}}\rangle)$, where the applicability predicate $\divT$ holds whenever the next instruction is a division instruction, and the leakage function leaks the value of register $\mathtt{RS1}$ and a corresponding leakage identifier.
    \item $A_3 = (\divT, \langle\mathtt{``Reg[RS2]"}, \myvalue{\mathtt{Reg[RS2]}}\rangle)$ similarly leaks the value of the second source register $\mathtt{RS2}$ of division instructions.
\end{compactitem}
For the simple processor from \Cref{example:processor}, atom $A_3$ alone would be sufficient to capture its microarchitectural leakage, as neither the immediate of $\liT$ nor the first source operand of $\divT$ have any influence on its timing.
\end{example}

\subsubsection{Contracts and Contract Traces}
A set of contract atoms $S$ induces a \emph{contract}~$\contract_S$, which is a function from architectural states to sets of contract observations, \ie $\contract_S : \archStates \to \powerset{\ctrObs}$.  %
$\contract_S$ is defined by evaluating each applicable atom in $S$ as follows:
\[\contract_S(\sigma) := \{ \phi_A(\sigma) \mid  (\pi_A, \phi_A) \in S \wedge \pi_A(\sigma) \}\]
Given a sequence $\tau:=\sigma_0, \sigma_1, \ldots$ of architectural states, $\contract_S(\tau)$ denotes the corresponding sequence of observations $\contract_S(\sigma_0), \contract_S(\sigma_1), \ldots$.
Therefore, given an architectural state~$\sigma \in \archStates$, $\contract_S(\archEval(\sigma))$ denotes the \emph{contract trace}, \ie the sequence of contract observations associated with the execution $\archEval(\sigma)$.

\begin{example}[Contract trace]
Consider the following program:\\
    \[
        \begin{array}{l}
        \liT\ \mathtt{R1}, 6 \\ 
        \liT\ \mathtt{R2}, 7 \\ 
        \divT\ \mathtt{R3}, \mathtt{R1}, \mathtt{R2} \\
        \end{array}
    \]
    Let $S = \{A_1, A_2, A_3\}$ consist of the three contract atoms from \Cref{example:atoms}.
    Then, $\contract_S(\archEval(\sigma)) = (\{\langle ``\mathtt{imm}", 6\rangle\}, \{\langle ``\mathtt{imm}", 7\rangle\}, \{\langle ``\mathtt{Reg[RS1]}", 6\rangle, \langle ``\mathtt{Reg[RS2]}", 7\rangle\})$ for any initial architectural state $\sigma$ corresponding to the program.
\end{example}

Given a contract $\contract_S$, architectural states $\sigma$ and $\sigma'$ are \emph{contract indistinguishable} if $\contract_S(\archEval(\sigma)) = \contract_S(\archEval(\sigma'))$ and \emph{contract distinguishable}, otherwise.
For a contract to be satisfied, contract-indistinguishable states must also be indistinguishable to an attacker at the microarchitectural level, which we formalize next.

As defined above, contracts are specified on top of the architectural execution, which executes instructions in-order and non-speculatively.
While it is possible to capture microarchitectural leakage due to speculative execution in this framework\footnote{After all, the entire architectural state is available to the contract.}, it may be inconvenient and unintuitive to do so.
To capture speculative leakage more naturally, prior work~\cite{spectector2020,contracts2021} has shown how to explicitly model the effects of speculation at architecture level by following misspeculated instructions in the architectural execution.

Bloem et al.~\cite{power-contracts} have extended Sail with leakage statements that allow to specify contracts on top of the architectural execution.

In order to formalize the central notion of contract satisfaction,
we first need to model the abilities of an attacker at the microarchitectural level.

\subsubsection{Attackers}

Different attacker models have been proposed in the literature~\cite{Doychev2015} corresponding to different types of attacks and attacker abilities.
\emph{Time-based attackers}~\cite{DBLP:conf/crypto/Kocher96,bernstein_aes_attack,Aciicmez07,Tsunoo03} can measure the execution time of a program, which may even be possible remotely~\cite{Aciicmez07}.
\emph{Access-based attackers}~\cite{Percival:prime+probe+l1d,Osvik:prime+probe+l1d,exectimeexploit} can interact with the microarchitectural state, typically by timing their own memory accesses, which requires attacker and victim to share the same hardware platform, which is common in the cloud~\cite{Ristenpart:prime+probe+llc}.
\emph{Trace-based attackers}~\cite{Aciicmez06,Aldaya2019,Yarom14} monitor the victim's execution continuously obtaining information about the victim's activity across time, \eg by observing the timing of instruction retirements.\looseness=-1

Like prior work on leakage contracts~\cite{Wang23,Mohr24,wang2025synthesis}, in this formalization, we adopt a trace-based attacker model in which the attacker can extract information from the microarchitectural state in every cycle.
However, we note that other attacker models could equally be adopted.
Formally, we model a microarchitectural attacker as a function $\atk: \implStates \to \atkObs$ that maps microarchitectural states to attacker observations in $\atkObs$.
Given a microarchitectural state~$\sigma \in \implStates$, $\atk(\implEval(\sigma))$ denotes the \emph{attacker trace}, \ie the sequence of attacker observations associated with the cycle-accurate microarchitectural execution $\implEval(\sigma)$.
A pair of states $\sigma$ and $\sigma'$ is \emph{attacker distinguishable} if the corresponding attacker traces are different, \ie $\atk(\implEval(\sigma)) \neq \atk(\implEval(\sigma'))$.
Otherwise, they are \emph{attacker indistinguishable}.\looseness=-1

\subsubsection{Contract Satisfaction}
A microarchitecture $\uarch$ \emph{satisfies} a contract $\contract_S$ (equivalently, $\contract_S$ is \emph{sound} for $\uarch$) with respect to attacker $\atk$ if the attacker cannot infer more about the initial architectural state by observing $\uarch$’s executions than what is revealed through the contract’s observations. Concretely, for any two initial states that coincide in their microarchitectural components\footnote{Differences in the initial microarchitectural state may lead to attacker distinguishability that is independent of the architectural execution.}, if the contract traces are identical, then the attacker traces must also be identical—i.e., the attacker cannot distinguish between the two executions.
This is formalized in \Cref{def:contract-satisfaction}.\looseness=-1

\begin{definition}[Contract satisfaction~\cite{contracts2021}]\label{def:contract-satisfaction}
Microarchitecture $\uarch$ \emph{satisfies} contract $\contract_S$ for attacker $\atk$, denoted by $\ctrsat{ \contract_S }{ \atk }$, if for all states $\sigma,\sigma' \in \implStates$ 
    with  \[\uarchProj{\sigma} = \uarchProj{\sigma'} \text{ and }\contract_S(\archEval(\archProj{\sigma})) = \contract_S(\archEval(\archProj{\sigma}'))\] 
    we have \[\atk(\implEval(\sigma)) = \atk(\implEval(\sigma')).\]
\end{definition}

\begin{example}[Contract satisfaction]\label{example:satisfaction}
Consider the architecture $\arch$ and the microarchitecture $\uarch$ from \Crefrange{example:isa}{example:processor} and the contract $\contract_S$ defined by the singleton set of atoms $S = \{A_3\}$, where $A_3 = (\divT, \langle\mathtt{``Reg[RS2]"}, \myvalue{\mathtt{Reg[RS2]}}\rangle)$ as in \Cref{example:atoms}.
Let the attacker $\atk$ expose the timing of instruction retirements, \ie in every cycle the attacker can see whether or not an instruction retires.

Then we have $\ctrsat{ \contract_S }{ \atk }$, \ie the contract exposes at least as much information about the execution as the attacker can infer from its observations on the microarchitecture.

In fact, in this case, the contract exposes strictly more information:
While the microarchitectural attacker can only determine whether the divisor is $0$ or $1$, the contract exposes the entire divisor.
We can strengthen the contract to better reflect the actual leakage of $\uarch$ by replacing atom $A_3$ with atom $A_3' = (\divT, \langle\mathtt{``Reg[RS2]"}, \myvalue{\mathtt{Reg[RS2]} \in \{0, 1\}}\rangle)$. Atom $A_3$ only reveals whether the divisor is $0$ or $1$.
\end{example}

\begin{example}[Contract non-satisfaction]
Consider the same architecture, microarchitecture, and attacker from \Cref{example:satisfaction} and a contract $\contract_T$ defined by the set of atoms $T = \{A_1, A_2\}$ with $A_1$ and $A_2$ from \Cref{example:atoms}.
Then $\ctrnonsat{ \contract_T }{ \atk }$, as demonstrated by the following program:
    \[
        \begin{array}{l}
        \divT\ \mathtt{R3}, \mathtt{R1}, \mathtt{R2} \\
        \end{array}
    \]
Consider two initial microarchitectural states $\sigma, \sigma'$ with $\sigma(\mathtt{R1}) = \sigma'(\mathtt{R1})= 0$, $\sigma(\mathtt{R2}) = 0$, and $\sigma'(\mathtt{R2}) = 42$.
Then, the execute stage will take 32 cycles from $\sigma'$ and only one cycle from $\sigma$, which means that the instruction retirements happen in different cycles and thus the two executions are attacker distinguishable. 
On the other hand, the only applicable contract atom exposes the value of register $\mathtt{R1}$, which is the same in both initial states and thus the executions are contract indistinguishable.
\end{example}

Contracts can be ordered by the security guarantees they provide to software. Intuitively, one contract is stronger than another if it ensures that less information is leaked to an attacker.
For example, the strongest contract expressible in the present formalism is defined by the empty set of atoms.
\begin{definition}[Preorder on contracts]
    Contract $\contract_A$ is stronger than $\contract_B$, denoted $\contract_B \sqsubseteq \contract_A$, if for all architectural states $\sigma, \sigma' \in \archStates$ with
    $\contract_B(\sigma) = \contract_B(\sigma')$
    we have 
    $\contract_A(\sigma) = \contract_A(\sigma')$.
\end{definition}
By construction, a smaller set of contract atoms induces a stronger contract: if $S \subseteq T$, then $\contract_T \sqsubseteq \contract_S$.
However, the converse does not hold in general, as different atoms may overlap in the information they expose.

As expected, a microarchitecture that satisfies a particular contract also satisfies all weaker contracts:
\begin{proposition}[Monotonicity of contract satisfaction]\label{prop:monotonicity}
    If $\ctrsat{ \contract_A }{ \atk }$ and $\contract_B \sqsubseteq \contract_A$, then $\ctrsat{ \contract_B }{ \atk }$.
\end{proposition}

Contract satisfaction is tightly connected to functional correctness as mentioned earlier in \Cref{sec:functionalcorrectness}.
Imagine a microarchitecture that does not correctly implement the architecture, \ie the microarchitectural execution sometimes deviates from the prescribed architectural execution.
If this deviation may lead to incorrect control flow, it is virtually impossible to adhere to a non-trivial contract, as the contract leakage is defined on top of the architectural execution.

To separate functional and security concerns, Wang et al.~\cite{Wang23} have shown how to decouple the two concerns by introducing the notion of \emph{microarchitectural contract satisfaction}.
The observation underlying this notion is that functional correctness implies that correct architectural traces can be recovered from microarchitectural traces.
Thus, microarchitectural contract satisfaction is defined directly on top of microarchitectural executions, without referring to the architecture:

\begin{definition}[Microarchitectural contract satisfaction~\cite{Wang23}]\label{def:microarchitectural-contract-satisfaction}
Microarchitecture $\uarch$ \emph{microarchitecturally satisfies} contract $\contract_S$ for attacker $\atk$ w.r.t. retirement predicate $\phi$, denoted by $\mctrsat{ \contract_S }{ \atk
}$, if for all states $\sigma,\sigma' \in \implStates$
    with  \[\uarchProj{\sigma} = \uarchProj{\sigma'} \text{ and }\contract_S(\archProj{\implEvalFilter{\phi}(\sigma)}) = \contract_S(\archProj{\implEvalFilter{\phi}(\sigma')})\] 
    we have \[\atk(\implEval(\sigma)) = \atk(\implEval(\sigma')).\]
\end{definition}

It is fairly easy to show that contract satisfaction and microarchitectural contract satisfaction are equivalent provided that the microarchitecture correctly implements the architecture:
\begin{theorem}[Decoupling~\cite{Wang23}]\label{thm:decoupling}
    If $\isasat{\uarch}{\arch}{\phi}$, then $\ctrsat{ \contract_S }{ \atk }$ if and only if $\mctrsat{ \contract_S }{ \atk }$.
\end{theorem}

Figure~\ref{fig:overview} illustrates Theorem~\ref{thm:decoupling} graphically.
Contract satisfaction straddles both functional concerns (separated vertically) and security concerns (separated horizontally), but can be decoupled into the two orthogonal notions of ISA compliance and microarchitectural contract satisfaction.

\newcommand{\exatk}{\atk}
\newcommand{\exarch}{\arch}
\newcommand{\exuarch}{\uarch}
\newcommand{\excontract}{excontract}

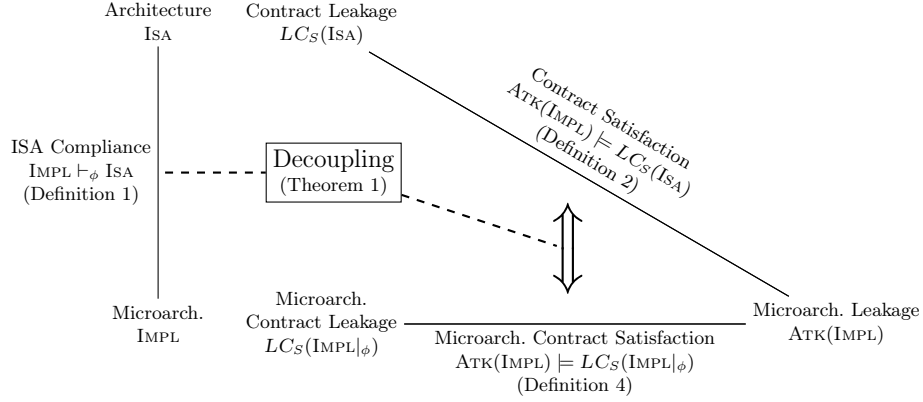
\begin{figure*}[t]
\centering
\small
\begin{tikzpicture}[scale=0.8, transform shape,every text node part/.style={align=center}]
    \node (topleft) at (13,0) {Microarch. Leakage\\
    $\compose{\exatk}{\exuarch}$};
    \node (topright) at (4.5,0) {Microarch.\\Contract Leakage\\
    $\compose{\contract_S}{\uarch|_{\phi}}$};
    \node (bottomright) at (4.5,4.95) {Contract Leakage\\
    $\compose{\contract_S}{\exarch}$};
    \node (topleftleft) at (1.8,0) {Microarch.\\$\exuarch$};
    \node (bottomleftleft) at (1.8,5) {Architecture\\$\exarch$};

    \node[draw] (decoupling) at (4.7,2.5) {\large Decoupling\\(\Cref{thm:decoupling})};

    \node (left) at (1.8,2.5) {};
    \node (right) at (8.5,1.25) {};

    \path[<->] (8.9,0.1) -- node[above,sloped] {\Huge $\Longleftrightarrow$} (8.9,2.4);

    \draw[dashed, thick] (left) -- (decoupling);
    \draw[dashed, thick] (decoupling) -- (right);
    
    \draw (topleft) -- (topright);
    \draw (topleft) -- (bottomright);
    
    \draw (topleft) -- node[below, sloped] {Microarch. Contract Satisfaction\\$\mctrsat{ \contract_S }{ \atk
}$\\(\Cref{def:microarchitectural-contract-satisfaction})} (topright);
    \draw (topleftleft) -- node[left] {ISA Compliance\\$\isasat{\exuarch}{\exarch}{\phi}$\\(\Cref{def:isa-compliance})} (bottomleftleft);
    \draw (topleft) -- node[above, sloped] {Contract Satisfaction\\$\ctrsat{ \contract_S }{ \atk }$\\(\Cref{def:contract-satisfaction})} (bottomright);
\end{tikzpicture}
\vspace{-7pt}
\caption{ISA compliance, contract satisfaction, microarchitectural contract satisfaction, and their relation via the decoupling theorem (adapted from \cite{Wang23}).\label{fig:overview}}
\end{figure*}

\subsubsection{Verifying Contract Satisfaction}
There is a growing body of work on verifying side-channel leakage properties of open-source processors~\cite{Wang23,shadow-logic,power-contracts}.
Some of this work is explicitly framed as verifying (microarchitectural) contract satisfaction~\cite{Wang23,shadow-logic,power-contracts}, while other work~\cite{conjunct,h-houdini} can be interpreted in that way.

LeaVe~\cite{Wang23} verifies microarchitectural contract satisfaction of DarkRISCV~\cite{darkriscv}, Sodor-2~\cite{sodor}, and Ibex~\cite{ibex}, relying on inductive invariants synthesized using the Houdini algorithm~\cite{flanagan2001houdini}.

Similarly, Contract Shadow Logic~\cite{shadow-logic} verifies microarchitectural contract satisfaction of \mbox{Sodor-2}~\cite{sodor} and a simple out-of-order processor using explicit-state model checking.
Work on power contracts~\cite{power-contracts} verifies security of the Ibex core against power-side-channel attacks using an inductive argument based on a manually defined mapping from microarchitectural to contract states.

Existing work on contract verification applies to relatively simple processors due to a lack of scalability of the underlying verification techniques~\cite{Wang23,shadow-logic,power-contracts} and the need for manual annotation~\cite{power-contracts}.
Application of these techniques to complex commercial cores will only be possible with further algorithmic improvements.
An alternative to improving scalability may be to adopt a contract-aware design methodology, where contract satisfaction follows by construction.

\subsubsection{Synthesizing Leakage Contracts}
Rather than merely verifying user-provided contracts, some recent work~\cite{wang2025synthesis,conjunct,h-houdini,Mohr24,rtl2mupath,rtl2muspec} aims to synthesize software-level characterizations of the microarchitectural leakage.
ConjunCT~\cite{conjunct} and VeloCT~\cite{h-houdini} determine a set of safe instructions that can be invoked with secret arguments without timing leaks.
Unfortunately, this rules out the use of branch and memory instructions, even on public inputs, which limits the utility of the characterization.
Notably, VeloCT~\cite{h-houdini} scales to more complex cores, such as the BOOM~\cite{Celio:EECS-2015-167} by deriving a set of relatively inductive properties.
It remains to be seen whether this approach can be applied to more expressive and fine-grained leakage contracts.
The focus of Mohr et al.~\cite{Mohr24} is to synthesize precise contracts, \ie contracts that distinguish as few attacker-indistinguishable executions as possible.
To this end, they rely on a contract template, spanning a large set of possible leakage contracts, and a large set of test cases generated automatically.
While their work scales to the CVA6~\cite{cva6}, being based on observing a finite set of executions, it is not guaranteed to synthesize sound contracts.
Subsequent work on LeaSyn~\cite{wang2025synthesis} combines LeaVe~\cite{Wang23} with the approach by Mohr et al.~\cite{Mohr24} to synthesize contracts that are provably sound and precise.
RTL2muPath~\cite{rtl2mupath,rtl2muspec} synthesizes \emph{microarchitectural execution paths}, which characterize how instructions may traverse a pipelined processor.
From these paths and \emph{leakage signatures}, which characterize how instruction operands may induce path variability, it is possible to derive leakage contracts in the sense of this tutorial paper.
In their experimental evaluation, Wang et al.~\cite{wang2025synthesis} determine that contracts synthesized in this way are fairly imprecise due to the coarse-grained underlying contract template.
While LeaSyn~\cite{wang2025synthesis} is able to synthesize more precise contracts, it relies on human intuition to define contract atoms that are able to capture the existing microarchitectural leakage well.
One challenge for future work is to automate this process.

\subsubsection{Contracts for Commercial Processors}
To support secure implementations of cryptographic algorithms, ARM and Intel recently introduced \emph{Data-Independent Timing} (DIT) instructions~\cite{ARMDIT} and \emph{Data-Operand Independent Timing} (DOIT) instructions~\cite{intelDOIT} to their ISAs.
Similarly, RISC-V has proposed the \emph{Zkt}~\cite{RISCVZKT} and \emph{Zvkt}~\cite{RISCVZKVT} extensions that attest that certain RISC-V instructions have data-independent timing.
ARM's DIT and Intel's DOIT require setting a mode bit to activate it, while processors supporting RISC-V's Zkt and Zvkt extensions always execute the respective instructions in constant time.
Both Intel's DOIT and RISC-V's Zkt/Zvkt extensions guarantee data-independent timing for the instructions' data operands, but their timing may still depend on other factors, ``such as addresses, immediate values, and instruction encoding''~\cite{TCHES:AOBGJLOS25}.

An interesting question is whether DIT, DOIT, Zkt, and Zvkt can be interpreted as leakage contracts in our framework.
Intuitively, their guarantees can be captured by exposing all of the operands of non-DIT/DOIT/Zkt/Zvkt instructions, while omitting the data operands of DIT/DOIT/Zkt/Zvkt instructions from the contract observations.
However, it is not as easy as that, as the ISAs do not preclude timing variability due to speculative execution.
Thus, it would at least be necessary to model speculation explicitly in the architectural execution, as done in prior work on speculation contracts~\cite{contracts2021}.
Precisely capturing the guarantees of DIT, DOIT, Zkt, and Zvkt as leakage contracts remains an open research question.

\subsubsection{Post-silicon Testing of Leakage Contracts}
Commercial processors are not amenable to formal analysis by academic researchers as their designs are closed source.
The complexity of these processors likely also precludes the in-house application of current contract verification algorithms by the designers themselves.
To scrutinize post-silicon commercial processors, researchers have resorted to systematic testing to uncover contract violations, which may subsequently lead to the discovery of vulnerabilities.
 
Scam-V~\cite{scamv} uses SMT solvers to synthesize a program together with a pair of initial inputs that are observationally equivalent, \ie contract indistinguishable in our terminology.
Then they execute the program with both inputs on actual hardware to determine whether they are attacker distinguishable, which would constitute a contract violation.
The focus in \cite{scamv} is on data-cache side channels and attacker distinguishability is determined by extracting and comparing the final cache states using ARM TrustZone.
Subsequent work~\cite{scamv-micro} extends \mbox{Scam-V}, introducing the notion of \emph{observation refinement}, which allows to focus the search for contract violations.
In this paper's terminology, they search for inputs that are contract indistinguishable under contract $A$, but distinguishable under a weaker contract $A+X$.
Choosing $X$ carefully allows to focus the search towards conjectured leakage.
This approach has led to the discovery of a vulnerability linked to speculative execution in the Cortex-A53.

Similarly, to Scam-V, Revizor~\cite{oleksenko2022revizor} aims to discover leakage in black-box processors in a contract-aware manner.
Revizor specifically targets speculation contracts~\cite{contracts2021} and it can automatically surface a range of speculative leaks, including Spectre, MDS, and LVI.
Subsequent work~\cite{oleksenko2023hide} showed how to improve %
fuzzing efficiency, by (a) filtering test cases that do not trigger microarchitecturally-visible speculation, and by (b) generating test cases that are by construction contract-indistinguishable.
This led to the discovery of several previously unknown speculative leaks in Intel and AMD processors.
Hofmann et al.~\cite{hofmann2023speculation} adapted the contract framework and contract-aware fuzzing to focus on the effects of exceptions. 
Starting with a simple model, refining it each time a new, unpredicted leak is found, they systematically document transient behaviors and discover three speculative vulnerabilities related to exceptions.

\section{Secure Programming}\label{sec:programming}

\newcommand{\erase}{\textit{erase}}

Secure programming aims to prevent confidential data from leaking to an attacker.
To capture this formally, we need to distinguish between public and private data.
Given a policy $\pi$ that decides which data is public and which data is private, let $\equiv_\pi$ be the induced equivalence relation on states, such that $\sigma \equiv_\pi \sigma'$ if $\sigma$ and $\sigma'$ differ only in private data\footnote{We cannot define $\equiv_\pi$ based on $\pi$ more formally as our framework currently makes no assumptions about the internal structure of architectural states.}.

If all executions of a program $P$ starting from policy-equivalent states are indistinguishable to the attacker, then the attacker cannot learn anything about private data.
This leads to the following definition:
\begin{definition}[Non-interference w.r.t.\ an attacker]\label{def:noninterference:attacker}
    A program $P$ is \emph{non-interferent} w.r.t.\ policy $\pi$ on microarchitecture $\uarch$ under attacker $\atk$, denoted by $\nonint$, if for all initial states $\sigma, \sigma' \in \implStates$ with \[\prog(\archProj{\sigma}) = \prog(\archProj{\sigma}') = P \text{ and } \archProj{\sigma} \equiv_\pi \archProj{\sigma}'\] we have \[\atk(\implEval(\sigma)) = \atk(\implEval(\sigma')).\]
\end{definition}

The above definition directly refers to the attacker on the microarchitecture.
For secure programming we would rather consider a program w.r.t. a contract, which is defined analogously:
\begin{definition}[Non-interference w.r.t. a contract]\label{def:noninterference:contract}
    A program $P$ is \emph{non-interferent} w.r.t.\ policy $\pi$ on contract $\contract_S$, denoted by \nonintcontract, if for all initial states $\sigma, \sigma' \in \archStates$ with \[\prog(\sigma) = \prog(\sigma') = P \text{ and } \sigma \equiv_\pi \sigma'\] we have \[\contract_S(\archEval(\sigma)) = \contract_S(\archEval(\sigma')).\]
\end{definition}

If we can guarantee that a program is non-interferent w.r.t. a contract and a microarchitectures satisfies this contract, then non-interference transfers to the microarchitecture:
\begin{proposition}[Non-interference~\cite{contracts2021}]
    If $\ctrsat{ \contract_S }{ \atk }$ and $\nonintcontract$, then $\nonint$. 
\end{proposition}
Thus, leakage contracts enable secure programming without explicitly considering the microarchitecture, which is the main benefit of the contract framework.

\begin{example}[Non-interference]
Consider the scenario from \Cref{example:satisfaction}, where we have $S = \{A_3\}$ and $\ctrsat{ \contract_S }{ \atk }$, with
\[    A_3 = (\divT, \langle\mathtt{``Reg[RS2]"}, \myvalue{\mathtt{Reg[RS2]}}\rangle)\]
and the following program:
    \[
        \begin{array}{l}
        \liT\ \mathtt{R2}, 7 \\ 
        \divT\ \mathtt{R3}, \mathtt{R1}, \mathtt{R2} \\
        \end{array}
    \]
In the first instruction, $\liT\ \mathtt{R2}, 7$, atom $A_3$ is not applicable and we get the empty set of observations.
In the second instruction, $\divT\ \mathtt{R3}, \mathtt{R1}, \mathtt{R2}$, atom $A_3$ is applicable, leaking the value of the second operand, but it is always $7$.
Thus, the program generates the same contract trace under any possible initial state, namely $\contract_S(\archEval(\sigma)) = (\{\}, \{\langle ``\mathtt{Reg[RS2]}", 7\rangle\})$.
Thus, the program is non-interferent under any policy and, due to contract satisfaction, it is also non-interferent w.r.t. to the attacker $\atk$.

On the other hand, the following program is only non-interferent w.r.t. $\contract_S$ under policies in which register $\mathtt{R2}$ is public: $\divT\ \mathtt{R3}, \mathtt{R1}, \mathtt{R2}$.
\end{example}

Two fundamentally different types of secure programming have been studied in the literature: \emph{sandboxing} and \emph{constant-time programming}.
Sandboxing enables the safe execution of untrusted code, preventing it from accessing secret data out of its designated memory region, i.e., the sandbox.
Constant-time programming, on the other hand, enables computing on secret data, like cryptographic keys, while preventing the leakage of this secret data to side-channel attackers. 
Both paradigms can be captured as non-interference w.r.t. appropriate policies and contracts.
In constant-time programming, the policy classifies secret data as private and public data as public, while in sandboxing, the policy classifies data inside the sandbox as public and data outside the sandbox as private.\looseness=-1

\newcommand{\ctverif}{\textsc{ct-verif}\xspace}

\subsubsection{State-of-the-art in Secure Programming}
The emergence of timing attacks~\cite{DBLP:conf/sp/AlFardanP13,bernstein_aes_attack,10.5555/1090583.1648610,exectimeexploit,DBLP:conf/crypto/Kocher96,Percival:prime+probe+l1d} on cryptographic libraries triggered the development of \emph{constant-time programming} disciplines~\cite{bernstein_aes_attack,DBLP:conf/ccs/GarciaBY16,DBLP:conf/icisc/MolnarPSW05,AlmeidaBBDE16}, software-level rules that aim to make timing attacks impossible. 
Such rules can be captured in leakage contracts.
A common pre-Spectre assumption was that code would be ``constant time'', \ie not leak secrets to side-channel adversaries, if it (1) did not branch on secrets, (2) did not perform secret-dependent memory accesses, and (3) did not use secrets as operands of variable-time instructions.
These assumptions can be captured via a leakage contract consisting of atoms that expose the control flow, memory access patterns, and operand values of variable-time instructions.

The \ctverif tool~\cite{AlmeidaBBDE16} analyzes LLVM assembly code w.r.t. a leakage model that encodes these rules.
It relies on a self-composition~\cite{BartheDR11} approach, where two copies of a program are composed to analyze executions from policy-equivalent states.
\ctverif employs \textsc{SMACK}~\cite{DBLP:conf/cav/RakamaricE14} to translate the LLVM code to the Boogie intermediate verifiation language (IVL)~\cite{leino2008this}, which in turn applies SMT solvers to discharge verification conditions.
Blazy et al.~\cite{BlazyPT19} analyzes constant-time implementations at the source level applying abstract interpretation to an instrumented version of the program that tracks information flow and prevents branching on secrets.

Speculative-execution attacks~\cite{spectre2019} demonstrate that the leakage contract underlying common constant-time assumptions was not satisfied by modern processors. 
In response, researchers introduced the notions of \emph{speculative non-interference}~\cite{spectector2020} and \emph{speculative constant-time}~\cite{DBLP:conf/pldi/CauligiDGTSRB20} to capture potential information leakage due to speculative execution.
Based on these notions, they introduced \textsc{Spectector}~\cite{spectector2020,DBLP:conf/ccs/FabianGP22}, \textsc{Pitchfork}~\cite{DBLP:conf/pldi/CauligiDGTSRB20}, and \textsc{binsec/rel}~\cite{daniel2020binsec} analyzers that detect constant-time violations using symbolic execution.
Subsequent work~\cite{contracts2021} analyzed the ISA-level security guarantees provided by secure speculation mechanisms and formalized their findings using hardware-software leakage contracts, which may form the basis for software-level analysis of vulnerabilities.
Arranz-Olmos et al.~\cite{TCHES:AOBGJLOS25} study the use of Intel's DOIT to securely implement cryptographic algorithms taking into account the effects of speculative execution.

Existing analysis tools~\cite{AlmeidaBBDE16,spectector2020,daniel2020binsec,DBLP:conf/pldi/CauligiDGTSRB20} are based on a hard-coded leakage contract.
As new vulnerabilities are discovered and fine-grained leakage contracts are synthesized for open-source processors~\cite{wang2025synthesis,conjunct,h-houdini,Mohr24,rtl2mupath,rtl2muspec} a challenge for future work is to develop analyses that are easily adaptable to new contracts, \eg by taking its definition as an input.
It is conceivable to employ a binary-level analysis platform like \textsc{Binsec}~\cite{binsec} for this purpose.
Analyses based on symbolic execution are limited in scalability.
Developing more scalable analyses, \eg based on abstract interpretation~\cite{BlazyPT19}, is an important challenge for future work.\looseness=-1

Leakage contracts, as defined in this paper, apply at the ISA level.
There is a growing body of work on preservation and enforcement of non-interference properties during compilation~\cite{patrignani2021exorcising,DBLP:conf/sp/CauligiDMBS22,blade,DBLP:journals/pacmpl/WallM25,Barthe19constanttimecompiler,DBLP:journals/pacmpl/OlmosBBGL25,DBLP:conf/csfw/BartheGL18}, which require corresponding notions at the source and target levels.

CHERI~\cite{DBLP:conf/isca/WoodruffWCMADLNNR14} is an ISA that replaces raw pointers with unforgeable, hardware-enforced capabilities.
These capabilities carry bounds and permissions.
Memory accesses are checked against these capabilities, enabling strong memory safety and fine-grained compartmentalization.
Thus, CHERI promises to enable efficient sandboxing of untrusted code.
However, CHERI does not directly address side-channel security:
The Spectre attacks~\cite{spectre2019} demonstrated that speculative execution can be exploited to bypass architectural sandboxing mechanisms such as the ones enforced by CHERI.
An interesting question explored in recent work~\cite{DBLP:conf/iccd/FuchsWRJ0BDNWM24} is how to efficiently implement speculation while preserving CHERI's memory safety guarantees even in the presence of side-channel adversaries.

Out of the box, CHERI appears less suited for constant-time programming.
E.g. it is perfectly possible to write non-constant-time CHERI code that leaks information through timing side channels by branching on secret data or performing secret-dependent memory accesses, as long as these accesses are within the bounds of the capabilities.
Recent work~\cite{DBLP:conf/ccs/ElAtaliGNA25} explores extending CHERI with \emph{blinded capabilities} that inform the hardware about secret data. 
Correct implementations of blinded capabilities prevent timing leaks due to secret-dependent control flow and memory accesses.

\section{Concluding Remarks}\label{sec:conclusions}

In this tutorial paper, we have introduced the foundations of hardware-software leakage contracts.
Leakage contracts act as a bridge between hardware and software: 
they serve as a foundation for secure programming and as a specification for the development of the underlying microarchitectures.
There are still significant open challenges in this area, which we have highlighted throughout the paper, in particular on the hardware side:
How can we design complex high-performance processors that provably satisfy useful leakage contracts?
Current verification techniques do not scale to complex commercial processors.
At the same time, we lack contract-aware hardware design methodologies.

\begin{credits}
\subsubsection{\ackname} 
I would like to thank my collaborators on leakage contracts, in particular Marco Guarnieri, Boris Köpf, Zilong Wang, Klaus v. Gleissenthall, and Gideon Mohr for their contributions to this research direction, as well as the anonymous reviewers for their helpful comments on earlier versions of this paper.
This work has received funding from the European Research Council under the European Union's Horizon 2020 research and innovation programme (grant agreement no. {101020415}).
\end{credits}
\bibliographystyle{splncs04}
\bibliography{biblio}
\end{document}